# Meter-scale spark X-ray spectrum statistics

B. E. Carlson[1,2], N. Østgaard[2], P. Kochkin[3], Ø. Grondahl[2], R. Nisi[2], K. Weber[1], Z. Scherrer[1], and K. LeCaptain[1]

[1]Department of Physics, Carthage College, Kenosha, Wisconsin, USA, [2]Birkeland Center for Space Science, University of Bergen, Bergen, Norway, [3]Technische Universiteit Eindhoven, Eindhoven, Netherlands

**Abstract** X-ray emission by sparks implies bremsstrahlung from a population of energetic electrons, but the details of this process remain a mystery. We present detailed statistical analysis of X-ray spectra detected by multiple detectors during sparks produced by 1 MV negative high-voltage pulses with 1 μs risetime. With over 900 shots, we statistically analyze the signals, assuming that the distribution of spark X-ray fluence behaves as a power law and that the energy spectrum of X-rays detectable after traversing ∼2 m of air and a thin aluminum shield is exponential. We then determine the parameters of those distributions by fitting cumulative distribution functions to the observations. The fit results match the observations very well if the mean of the exponential X-ray energy distribution is 86 ± 7 keV and the spark X-ray fluence power law distribution has index −1.29 ± 0.04 and spans at least 3 orders of magnitude in fluence.

## 1. Introduction

Spark discharge is a complicated multiscale process. The physics involved ranges from subnanosecond submillimeter electron avalanches governed by atomic and molecular ionization and attachment cross sections to centimeter-scale propagating space charge waves (streamers) to microsecond-scale meter-long channels of ionized gas described by nonequilibrium plasma and electrodynamic behavior. On a larger scale, spark behavior in nature ranges from 100 km lightning channels that last seconds and span conditions from the upper reaches of a thundercloud to the microsecond-scale processes by which such a channel attaches to the ground.

In such a diverse range of behavior one can find many puzzling phenomena. The process by which a discharge initiates in the cloud is not understood nor is the process by which lightning channels (leaders) extend and branch. The extension process sometimes proceeds as a series of microsecond-scale steps separated by 10 s of microseconds of relative quiescence. More puzzling, the steps seem to be immediately preceded by formation of a hot conductive channel displaced from the end of the existing channel that rapidly connects with the main channel, though the mechanism by which this "space leader" forms is not understood. On a smaller scale, the transition from streamer to leader is the focus of much research as are the dynamics of the streamer itself and the role of energetic particles in the process. See *Dwyer and Uman* [2014] for a review of recent results and open questions.

This role of energetic particles is especially interesting in the light of observations of X-ray production by lightning and the hypothesis that energetic particles produced by lightning contribute to the production of terrestrial gamma ray flashes. X-rays as produced by lightning seem to occur coincident with stepwise extensions of the channel [*Howard et al.*, 2010; *Dwyer et al.*, 2011] and have energies around a few 100 keV [*Moore et al.*, 2001; *Dwyer et al.*, 2003; *Dwyer*, 2005b]. These X-rays imply the existence of a much larger population of higher-energy electrons. While low-energy electrons encounter high dynamic friction, relativistic electrons encounter much lower friction and can "run away" to very high energies when driven by electric fields, a fact first recognized by *Wilson* [1925] and modeled in the context of electric fields near lightning channels in *Carlson* [2009, chapters 5 and 6]. "Seed" energetic electrons necessary to initiate such a process can come from energetic background radiation (i.e., cosmic rays [*Carlson et al.*, 2008]), low-energy electrons in local electric fields strong enough to overcome the maximum friction force [e.g., *Moss et al.*, 2006; *Li et al.*, 2009; *Chanrion and Neubert*, 2010], or by feedback from prior generations of energetic electrons [*Dwyer*, 2003, 2007]. Regardless, electric fields may drive avalanche growth of populations of such runaway electrons [*Gurevich et al.*, 1992].

Unfortunately, the relative importance of these processes for phenomena like X-ray production by lightning or terrestrial gamma ray flashes is not well understood.

In the context of these puzzles, any additional information about the processes involved may potentially be useful. While limited in scale, laboratory studies can shed light on the detailed dynamics of leader extension, and while the energy scales are necessarily smaller (of order 1 MV compared to the 10 s of megavolts of natural lightning), lab sparks even produce X-rays.

X-ray production by meter-scale sparks has been observed on several occasions. *Dwyer* [2005a] report the first detection of X-rays associated with such sparks in a study of 14 discharges of 1.5 MV, with total energy deposited in a detector up to MeV scale. Variation of signals among X-ray detectors with different attenuators suggest X-ray energies from 30 keV to 150 keV piling up to produce the MeV-scale observations. *Rahman et al.* [2008] report similar results with total energy deposited up to several MeV with 1 MV discharge. *Nguyen et al.* [2008] (elaborated in *Nguyen* [2012]) also report similar results for 0.88–1 MV sparks with a much larger data set and with a variety of detectors, attenuators, and positions of detector. *Nguyen et al.* [2008] show intensity variations with high-voltage (HV) polarity and with distance between detector and spark gap, suggesting that such intensity variations could result from X-ray production by positive streamers from metal structures near the detectors. This highlights the importance of positioning detectors at a sufficient distance from the gap. *Dwyer et al.* [2008] report a further study of 241 discharges of 1 MV with various polarities and gap distances and push the maximum energy deposited in a single detector up to 50 MeV with varied attenuators providing evidence for individual photon energies exceeding 300 keV and statistical evidence for an average photon energy at most 230 keV. *Dwyer et al.* [2008] also report collimator experiments supporting the production of X-rays by a diffuse source in the gap between the electrodes and that bursts of X-rays produced late in the discharge come from elsewhere. *March and Montanyà* [2010] report that HV pulses with rapid risetime tend to produce more X-rays, and *March and Montanyà* [2011] examines the effect of electrode geometry. *Kochkin et al.* [2012], *Kochkin et al.* [2014], and *Kochkin et al.* [2015] add nanosecond-resolution photography as a useful tool, mapping the streamer clouds produced by positive [*Kochkin et al.*, 2012] and negative [*Kochkin et al.*, 2015] HV discharges. In such experiments the HV risetime is much longer than the time for streamers to propagate across the gap, and in experiments with negative high voltage the negative streamers tend to appear and grow in bursts [see, e.g., *Kochkin et al.*, 2014, Figure 6]. *Kochkin et al.* [2015] demonstrate for negative HV that the X-rays tend to be emitted during these bursts of streamer development but are emitted on a much shorter timescale than the burst, supporting the suggestion that the large and transient electric fields that result from interaction of negative and positive streamer fronts may play a role in energetic electron and thus X-ray production. *Kochkin et al.* [2012, 2015] also describe that attenuator experiments they claim are consistent with 200 keV characteristic X-ray energy. These estimates are based simply on comparison of registration rate (the fraction of sparks for which X-rays are observed) for various attenuators with the expected attenuation in number of photons incident on the detector, assuming all sparks emit the same number of photons and that pileup does not affect their observations, though they acknowledge that pileup has a significant effect.

Together, these studies provide a reasonably complete picture of X-ray emissions from sparks: free electrons are pushed to overcome friction in the high-field region ahead of a negative streamer (possibly briefly enhanced by a nearby positive streamer), gain at least enough energy to run away in the lower fields surrounding the streamer, then undergo bremsstrahlung in the surrounding air to emit X-rays. These intense streamer fields exist only as the discharge develops and are unlikely to exist during the discharge itself, though the intense electrodynamic environment during discharge may drive discharges elsewhere in the lab that also produce detectable X-rays. However, this picture remains incomplete in several aspects: beyond an "average" energy, the properties of the photon spectrum are unknown and difficult to judge due to pileup, and the fact that some sparks produce copious X-rays while others produce no measurable X-rays is not addressed. The present paper attempts to shed light on these topics by examining the energies measured in X-ray detectors placed near a 1 m spark gap at the Technical University of Eindhoven during 921 shots as described in section 2. These shots give us the opportunity to statistically analyze the distributions of spark X-ray fluence, number of photons detected by multiple detectors, and photon energy. These distributions are described in section 3 and constrained by comparison to observations in section 4. We conclude and discuss the implications of our results in section 5.

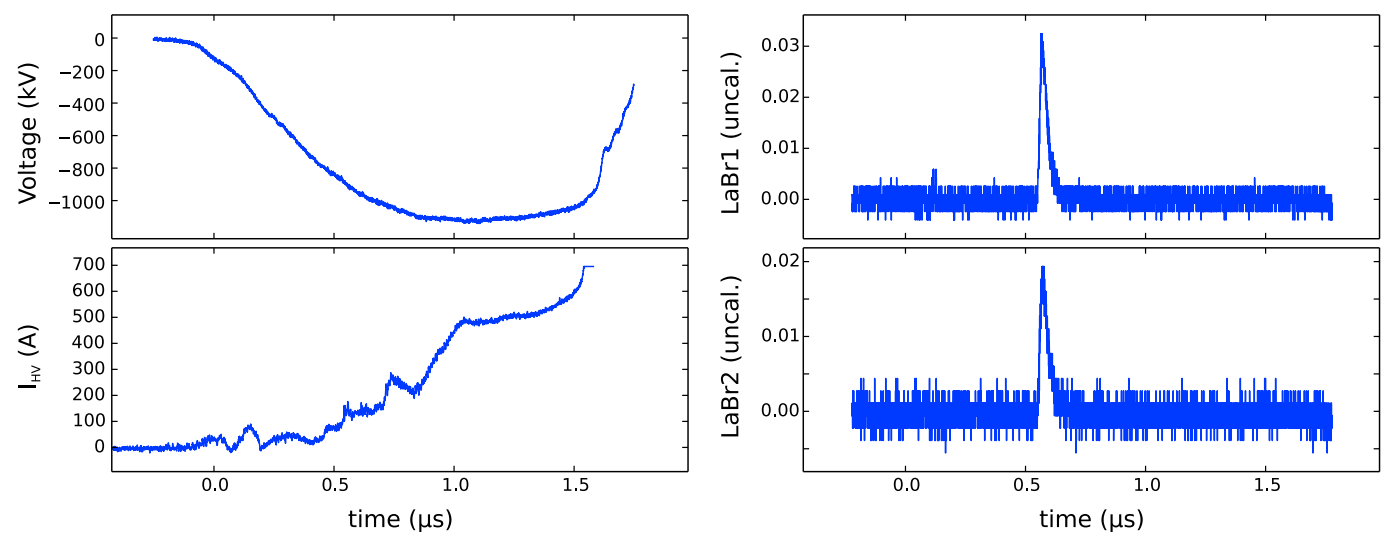

**Figure 1.** Sample records from a single spark. (top left) The voltage of the high-voltage electrode with respect to the ground electrode. (bottom left) The magnitude of the current flowing to the high-voltage electrode. (right) Signals from the two LaBr$_3$ detectors prior to calibration.

## 2. Experimental Setup

The experiment was conducted in the high-voltage lab at the Technical University of Eindhoven with a Haefly 2 MV Marx generator configured to produce 1 MV pulses with 1 μs risetime. All tests were carried out with a 1 m point-point gap with negative HV polarity. The spark voltage, ground electrode current, and high-voltage electrode current were recorded by Lecroy four-channel storage oscilloscopes configured to record 2 μs of data at 10 GHz sample frequency when triggered by the Marx generator, as were signals direct from the output of photomultiplier tubes monitoring a variety of energetic radiation detectors. No photomultiplier pulse shaping electronics was used. See *Kochkin et al.* [2012] for a more detailed description of the setup.

The radiation detectors used in the experiment included scintillating plastic optical fibers placed at a variety of locations around the spark to monitor energetic electrons. Analysis of these data has been presented [*Ostgaard et al.*, 2014] and will be published separately. Due to the need to run many shots with the scintillating fiber detectors at various locations, a total of 950 shots were carried out from which we recorded usable data for 921. During all shots, two additional X-ray detectors were running, providing a wealth of data about the photon population. These X-ray data are the focus of this paper.

These two detectors are each composed of a 1.5 inch long 1.5 inch diameter LaBr$_3$(Ce$^+$) scintillator monitored by a photomultiplier tube (PMT). These detectors are placed next to each other (separation ∼10 cm) in an electromagnetic compatibility (EMC) cabinet to shield the detectors from the electromagnetic noise produced by the spark. The location of the detectors relative to the spark corresponds to location H as shown in Figure 1 of *Kochkin et al.* [2015]. The scintillator material is separated from the spark by roughly 0.5 mm of aluminum in the scintillator housing and EMC cabinet wall and approximately 2 m of air.

The signals as recorded by the oscilloscopes are in volts and here have been converted to MeV by use of Cs-137 and Co-60 gamma ray emitters as sources of known energy. The lowest energy detectable by this setup is ∼20 keV, comparable to the minimum photon energies transmitted through the 0.5 mm aluminum shielding, while the maximum depends on the settings of the oscilloscope. Signal from detector 1 clips at just above 5 MeV, while detector 2 clips at just above 3.5 MeV.

Sample data are shown in Figure 1. The high-voltage trace shown is very consistent from one spark to the next, as is the high-voltage electrode current. The pulse-like features on the high-voltage electrode current occur when bursts of corona and streamer activity carry charge away from the electrode. See *Kochkin et al.* [2014] for a detailed discussion of such processes as seen in high-speed camera imagery. The timing of the X-ray pulses is also quite consistent. In this experiment, X-rays are typically not seen during the high-current phase of the discharge.

The pulses seen in the oscilloscope traces indicate deposition of energy in the scintillator, but a single pulse from the PMTs may be produced by multiple X-ray photons entering the scintillator. These photons may arrive at slightly different times and produce a visibly altered pulse shape, but typically no time structure is visible. The photon spectrum is therefore not directly measurable, and pulse pileup must be treated statistically.

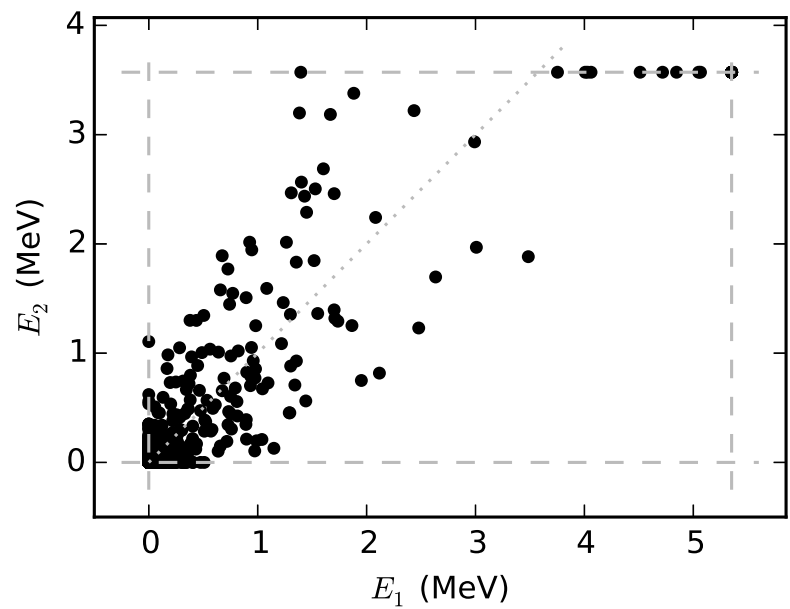

**Figure 2.** Scatterplot of observed energy deposition events. Each point represents a spark, where $E_1$ was deposited in detector 1, while $E_2$ was deposited in detector 2. The dashed lines show the 0 and saturation energy levels. The dotted line shows $E_1 = E_2$. At $E_1 = E_2 = 0$, 523 points overlap, and 20 points overlap at the upper right, where both detectors were saturated.

To compile data for such statistics, we search through the data for pulses. To search, we first smooth the data by low-pass filtering, then identify pulses as significant deviations from the background. Once pulses are identified, we collect a variety of data including pulse time, height, integral, and duration. Pulse height and integral are calibrated to energy as described above. Pulse integral is a more robust measure of deposited energy in case multiple pulses arrive near simultaneously, so our analysis here reports energy as determined by pulse integral. While pulse integral is somewhat resistant to saturation when oscilloscope signals clip, the energy of a saturated pulse cannot reliably be determined, so we enforce a maximum energy for each detector corresponding to the energy of the largest unclipped pulse. In cases when multiple pulses are observed in a single detector, we add their energies together to better capture variation from one spark to the next, though note that the majority of sparks have zero or only one burst so this combination of bursts only affects a small fraction of our data. The data set we consider here is therefore a set of pairs of numbers, each pair associated with a single spark and each number with the pair giving the total energy deposited in a detector during the given spark.

A scatterplot showing the energies deposited in each detector in each shot is shown in Figure 2. The energy deposited varies widely from one spark to the next. Most sparks (57%) produce no detectable signals in either detector, indicating less than 20 keV deposited, while approximately 3% of sparks saturate both detectors, indicating at least 3–5 MeV of energy deposited.

The points clearly cluster in the lower left, but a significant number of points appear in the middle and upper regions of the plot. Attempting to explain this distribution most simply, one might assume all sparks are identical and that all photon energies are equal and treat the observed distribution as solely due to Poisson statistical fluctuations in the number of photons observed. In this case, the high number of points in the lower left (57% of events undetected) implies a Poisson mean of 0.562. With this mean, observation of events that saturated both detectors is then incredibly unlikely as deposition of more than 8 MeV (3 MeV in one and 5 MeV in the other) requires at least 40 photons for the mean energy 200 keV consistent with the results quoted above. Observation of at least 40 photons from a Poisson distribution with mean 0.562 has a probability around $10^{-58}$, much less than the 3% of observed sparks that saturate both detectors.

Adding the complication of a photon energy distribution is logical, but does not help enough, since the maximum photon energy must be less than 1 MeV as the photons will be much less energetic than the 1 MV maximum spark voltage. Assuming the same Poisson mean as before and attempting to explain the events that saturated both detectors as due to an extreme fluctuation in photon energy such that all photons carried 1 MeV of energy means that we must have at least eight photons instead of at least 40, but the Poisson probability of observing at least eight photons given a mean of 0.562 is roughly $10^{-7}$, still much smaller than the 3% observed.

Clearly, there must be a significant variability in spark X-ray fluence at the location of the detectors from one spark to the next. This variability could come from intrinsic variability in spark X-ray luminosity or it could come from some variability in the geometry if X-ray emissions are not isotropic. Regardless, a full explanation of the distribution of points in Figure 2 must include the effect of the distribution of spark X-ray fluence, the distribution of numbers of photons detected by each detector (given the X-ray fluence), and the distribution of photon energies. These distributions and their properties are the focus of this paper.

## 3. Statistical Model

In attempting to build a statistical model of the distribution of points in Figure 2, we need to know the form of the distributions of physical properties relevant to the point locations. There are three main distributions at work: the distribution of spark X-ray fluence, the distribution of numbers of photons incident on each detector (given spark fluence), and the distribution of photon energies.

The distribution of spark fluence is determined in principle by the processes at work in electron acceleration. One spark has an intrinsically higher luminosity than another by having more energetic electrons, perhaps because random streamer branching events happened to lead to more negative streamers or perhaps fewer but more intense or more rapidly growing negative streamers or perhaps more interactions between negative streamers and positive streamers. Fluence variability may also result from anisotropy in the directionality of X-ray emissions: perhaps all sparks emit many X-rays, but the emissions are not always beamed toward the detectors. These processes may be chaotic or perhaps intrinsically random, but either way there will be some variability in X-ray fluence. To represent this variability, we treat the X-ray fluence as the expected value of the number of photons hitting a detector, $\eta$, and assume some probability distribution for the occurrence of various values of $\eta$. As there is yet no theoretical expectation for the form of this distribution, we must make some assumption. Examining the clustering of points in the lower left of Figure 2, we expect lower fluences to be more common than higher fluences, but the existence of points in the upper right suggests that the tail of the distribution to high fluences is quite strong. To represent distributions that follow this general trend of decreasing with a potentially long tail, we assume the distribution for $\eta$ is a power law with index $\lambda$:

$$p_\lambda(\eta) = \frac{(\lambda + 1)}{\eta_{max}^{\lambda+1} - \eta_{min}^{\lambda+1}} \eta^\lambda \qquad (1)$$

where $\eta_{min}$ and $\eta_{max}$ are lower and upper limits on $\eta$ necessary to ensure that the distribution is normalizable for $\lambda \geq -1$ and the leading constant ensures normalization. By examining the clustering in the lower left in Figure 2, we expect $\lambda < 0$, but the number of points saturating both detectors suggests that $\lambda$ is not too negative and that the distribution of $\eta$ is in some sense quite hard.

Given the fluence of the spark (i.e., given $\eta$), the number of photons hitting a detector, $N$, should be Poisson distributed with mean $\eta$. Normalized,

$$p_\eta(N) = \frac{\eta^N e^{-\eta}}{N!} \qquad (2)$$

This assumes that there is a very large number of photons produced by the spark, each with a very small probability of hitting the detector, and that the two detectors have the same probability of catching each photon. This implicitly assumes that X-ray emissions from the spark are uniform on the scale of the ∼10 cm separation of the detectors, but this is reasonable as the detectors are roughly 2 m from the spark and bremsstrahlung from electrons with only a few hundred keV of energy is not strongly beamed.

Finally, given a number of X-rays hitting each detector, one can calculate the distribution of deposited energy by applying a distribution of X-ray energies. This unfortunately is unknown as we do not know the distribution of electron energies responsible for the bremsstrahlung emissions [*Chanrion and Neubert*, 2010]. Figure 2 presents calculations of energy distributions of free electrons as produced in constant electric fields that appear linear on a semi–log plot (inset) and that show the effects of an exponential cutoff up to the threshold of runaway electron behavior. These exponential distributions of electron energy are characteristic of avalanche growth processes. The resulting photon distribution will share some of these characteristics, but in principle connecting the photon and electron distributions requires detailed treatment of bremsstrahlung, a topic beyond the scope of the current work. As such, we simply assume that the photon energy $\mathcal{E}$ is also exponentially distributed with mean photon energy $\mu$:

$$p_\mu(\mathcal{E}) = \frac{e^{-\mathcal{E}/\mu}}{\mu} \qquad (3)$$

where the division by $\mu$ ensures normalization for $0 < \mathcal{E} < \infty$. Note that for simplicity we are assuming that the photon energy distribution does not depend on X-ray fluence. It is reasonable to expect that $\mu$ depends on $\eta$ somewhat, but including this dependence would require still more assumptions and additional free parameters. Assuming an exponential photon energy distribution is especially convenient since the distribution of energies deposited in a detector $E$, a sum of the energies of $N$ photons ($E = \sum_{i=1}^{N} \mathcal{E}_i$) will be distributed as the sum of $N$ independent exponentially distributed random variables. The distribution of such a sum of exponential random variables can be expressed in closed form and is known as the Erlang distribution:

$$p_{N,\mu}(E) = \frac{E^{N-1} e^{-E/\mu}}{\mu^N (N-1)!} \tag{4}$$

Note, however, that if $N = 0$, no photons strike the detector, the energy deposited is exactly zero, and $p_{N=0,\mu}(E) = \delta(E)$, a Dirac delta function, instead of the continuous distribution above.

If the fluence of the spark ($\eta$) and the number of X-rays hitting each detector ($N_1$, $N_2$) were known, the joint distribution of energy depositions would simply be given by a product of two Erlang distributions. Unfortunately, the fluence of the spark and the number of X-rays hitting each detector are not known. As such, we must compute the marginal distribution of observed energies by integrating (i.e., taking an weighted average) over the possible values of the unknown quantities, weighted by the probability distributions associated with those quantities. For one detector,

$$p_{\lambda,\mu}(E) = \int_{\eta_{\min}}^{\eta_{\max}} d\eta \, p_\lambda(\eta) \sum_{N=0}^{\infty} p_\eta(N) p_{N,\mu}(E) \tag{5}$$

and for the joint distribution of simultaneous observations with two detectors,

$$p_{\lambda,\mu}(E_1, E_2) = \int_{\eta_{\min}}^{\eta_{\max}} d\eta \, p_\lambda(\eta) \sum_{N_1=0}^{\infty} \sum_{N_2=0}^{\infty} p_\eta(N_1) p_{N_1,\mu}(E_1) p_\eta(N_2) p_{N_2,\mu}(E_2) \tag{6}$$

where the integral over $\eta$ covers the range of possible spark X-ray fluence and the sums over $N$ or $N_1$ and $N_2$ cover the possible numbers of photons observed by each detector. The resulting joint probability distribution of $E_1$ and $E_2$ should be able to reproduce the observations in Figure 2.

Note that this is a hybrid continuous-discrete probability distribution. There is a finite probability that both detectors observe exactly zero energy, a discrete probability represented by the $N_1 = N_2 = 0$ term in the sums for which $p_{N_1,\mu}$ and $p_{N_2,\mu}$ become delta functions as described above. Likewise, the $N_1 = 0, N_2 > 0$, and $N_2 = 0, N_1 > 0$ terms in the probability distribution contain a single delta function. Only the terms for which both $N_1$ and $N_2$ are greater than zero lack delta function divergence behavior. These delta functions make the probability distribution difficult to work with directly but are capable of representing the clustering of points at the origin and along the axes in Figure 2.

These probability distributions form the basis of our model, and we seek to infer the parameters $\lambda$, $\mu$, $\eta_{\min}$, and $\eta_{\max}$ by comparison of our model to data.

## 4. Comparison to Data

Comparison of a probability distribution to data can be done in many ways, for example, the Kolmogorov-Smirnov test or the Anderson-Darling test. Such tests work on the basis of the cumulative distribution function (CDF) $F$, which for a probability distribution $p(x)$ in one dimension is defined as $F(x) = \int_{-\infty}^{x} p(\xi) d\xi$. Since here we work in two dimensions ($E_1, E_2$), we need a two-dimensional analog of the cumulative distribution function. Here we follow *Fasano and Franceschini* [1987] in taking our set of data points $\{P_i\} = \{(E_{1i}, E_{2i})\}$ and for each point, assigning four CDF-like values: $F_{\text{dat}\,i}^{<<}$, $F_{\text{dat}\,i}^{<\geq}$, $F_{\text{dat}\,i}^{\geq<}$, and $F_{\text{dat}\,i}^{\geq\geq}$, each giving the fraction of the observed data set in the corresponding quadrant relative to the data point in question. For example, $F_{\text{dat}\,i}^{<<}$ gives the fraction of data $\{P_j\}$ for which $E_{1j} < E_{1i}$ and $E_{2j} < E_{2i}$.

These $\{F_{\text{dat}\,i}^{\cdots}\}$ can be plotted if an order is assigned to the data points for use as an abscissa. Here it is convenient to order each set of $F_{\text{dat}\,i}^{\cdots}$ separately such that $F_{\text{dat}\,i}^{\cdots}$ is monotonically increasing, i.e., sort each set of $F_{\text{dat}\,i}^{\cdots}$ from smallest to largest and use its place in the resulting list as its abscissa. The order has no physical significance and is used only for plotting. The reordered $F_{\text{dat}\,i}^{\cdots}$ are shown in Figure 3. For example, the grey $\geq, \geq$ curve starts at 0.03 since 3% of sparks saturate both detectors: the data point that has the fewest data points with energies greater than or equal to its energy must be one of those doubly saturated events, and thus, 3% of the data satisfies the $\geq, \geq$ condition. Moving to the right along the $\geq, \geq$ curve, larger CDF values correspond to data points that are below and/or to the left of the upper right corner, i.e., data points for which larger fractions of the data satisfy the $\geq, \geq$ condition. The large jump from about 0.35 to 1.0 at CDF rank 400 occurs when incrementally increasing the fraction of the data set that satisfies the $\geq, \geq$ condition requires looking at a point where $E_1 = E_2 = 0$, which thus includes all other such points (57% of all data) as well as points for which $E_1 = 0$ or $E_2 = 0$. The CDF therefore jumps up to 1 since all of the data has $E_1 \geq 0$ and $E_2 \geq 0$. The direction of the inequality is related to the shape of the distribution, with $\geq\geq$ and $<<$ related to the lower left/upper right balance and width while the $<\geq$ and $\geq<$ versions related to upper left/lower right balance and width. One can imagine other ways of constructing such CDFs, but this is a straightforward technique that as we will see shortly is quite effective.

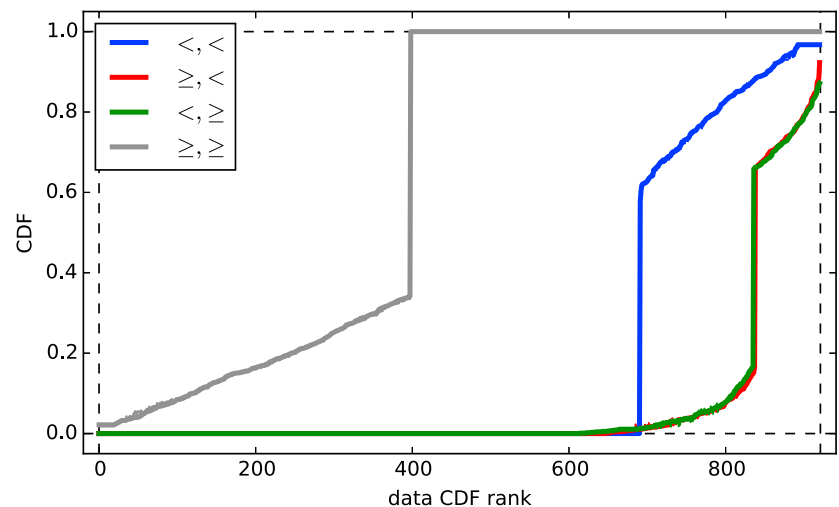

**Figure 3.** Cumulative distribution functions as calculated for the data. On a given curve, each horizontal point represents a point in the (2-D) data set, for which the vertical coordinate represents the fraction of the data set with energies related to the energies of the given data point by the inequalities associated with the given curve (i.e., the $F^{\cdots}_{\text{dat}\,i}$ described in the text). Each curve represents a different inequality, and the data have been reordered for each curve to make the curve in question monotonically increasing. Two sets of curves are shown, thick curves from the data and thin curves predicted given the best fit parameters discussed in the text, but the curves overlap so much any deviations are difficult to judge.

Such curves as in Figure 3 can be predicted on the basis of the probability density (equation (6)), assigning a predicted $F^{\cdots}_{\text{pred}\,i}$ to each data point by integrating the expected distributions over the region relevant to the inequalities in question. For example,

$$F^{<<}_{\text{pred}\,i} = \int_{-\infty}^{E_{1i}} dE_1 \int_{-\infty}^{E_{2i}} dE_2\, p_{\lambda,\mu}(E_1, E_2) \qquad (7)$$

Similar calculations can be made for predicted $F^{\geq<}_{\text{pred}\,i}, F^{<\geq}_{\text{pred}\,i}, F^{\geq\geq}_{\text{pred}\,i}$ by changing the limits on the integrals to go from the point in question to $+\infty$ as appropriate. Regardless, the exponentials and powers in the integrand can be manipulated to convert the integrals here and in equation (6) into upper and lower incomplete gamma function evaluations (e.g., $\Gamma(k, x) = \int_x^\infty x^{k-1} e^{-x} dx$), and the summations in equation (6) can be computed numerically and truncated without significant loss of accuracy.

For a given $\lambda$, $\mu$, $\eta_{\min}$, and $\eta_{\max}$, the result of this exercise is four sets of numbers ($\{F^{<<}_{\text{pred}\,i}\}$, $\{F^{\geq<}_{\text{pred}\,i}\}$, $\{F^{<\geq}_{\text{pred}\,i}\}$, and $\{F^{\geq\geq}_{\text{pred}\,i}\}$) that can be compared to the corresponding sets associated with the data ($\{F^{<<}_{\text{dat}\,i}\}$, $\{F^{\geq<}_{\text{dat}\,i}\}$, $\{F^{<\geq}_{\text{dat}\,i}\}$, and $\{F^{\geq\geq}_{\text{dat}\,i}\}$). Treating these numbers as analogous to the cumulative distribution function, the Kolmogorov-Smirnov test statistic would be the maximum deviation between any two corresponding $F^{\cdots}_{\cdots\,i}$. The Kolmogorov-Smirnov test is sometimes criticized as not very powerful [e.g., *Razali and Wah*, 2011], so instead, we calculate an analog of the Anderson-Darling test statistic:

$$S = \sum_i \left( \frac{(F^{<<}_{\text{dat}\,i} - F^{<<}_{\text{pred}\,i})^2}{F^{<<}_{\text{pred}\,i}(1 - F^{<<}_{\text{pred}\,i})} + \frac{(F^{\geq<}_{\text{dat}\,i} - F^{\geq<}_{\text{pred}\,i})^2}{F^{\geq<}_{\text{pred}\,i}(1 - F^{\geq<}_{\text{pred}\,i})} + \frac{(F^{<\geq}_{\text{dat}\,i} - F^{<\geq}_{\text{pred}\,i})^2}{F^{<\geq}_{\text{pred}\,i}(1 - F^{<\geq}_{\text{pred}\,i})} + \frac{(F^{\geq\geq}_{\text{dat}\,i} - F^{\geq\geq}_{\text{pred}\,i})^2}{F^{\geq\geq}_{\text{pred}\,i}(1 - F^{\geq\geq}_{\text{pred}\,i})} \right) \qquad (8)$$

i.e., the squared deviation between the predicted CDF and the observed CDF, weighted as in the Anderson-Darling statistic, summed over all four CDF types (inequalities), and summed over all points in the observed data set. Some numerical problems arise when both numerator and denominator are exactly zero, i.e., one of the $F^{\cdots}_{\text{pred}\,i}$ is exactly 0 or 1, as occurs, for example, for $F^{<<}_{\text{pred}\,i}$ for a point, where $E_{1i} = E_{2i} = 0$. Since such points involve exact match between predicted and observed CDF and therefore should not contribute to the test statistic, we add a small factor (0.001) to the denominator of each term to ensure the division of zero by zero results in zero. This does not significantly affect the other terms in the test statistic or the overall results.

We then fit our predicted distribution to the observed distribution—minimizing $S$—by varying the parameters of our calculated CDF with the downhill simplex algorithm (*Nelder and Mead* [1965] as implemented in *Johnson* [2014]). The optimal values are $\lambda = -1.29$, $\mu = 86$ keV, $\eta_{\min} = 0.022$, and $\eta_{\max} = 113$.

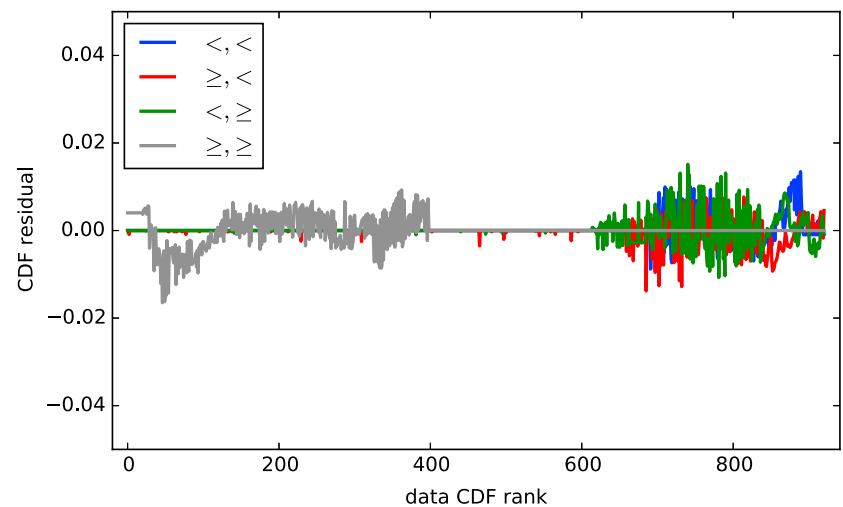

**Figure 4.** Cumulative distribution function residuals for the best fit results as discussed in the text. A positive deviation indicates a larger data CDF than computed from the best fit parameters. Deviations are typically less than 1%.

Since our multidimensional CDF and modified Anderson-Darling test statistic are so far removed from their original application, we only use the statistic in judging the quality of the fit in the optimization process described above and make no attempt to apply the distributions typically associated with the Anderson-Darling test statistic to assess the uncertainty in our fit results. Instead, we judge uncertainty by bootstrap, repeating our fit process many times with alternative data sets produced from our original data by sampling with replacement. The fit results are each approximately normally distributed, with mean and standard deviation given as follows: $\lambda = -1.29 \pm 0.04$, $\mu = 86 \pm 7$ keV, $\eta_{min} = 0.022 \pm 0.006$, and $\eta_{max} = 110 \pm 25$. Plotting these bootstrapped fit results, one parameter versus another, shows very little correlation between results, with the exception of $\lambda$ and $\eta_{min}$ which are negatively correlated: more negative $\lambda$ is associated with higher $\eta_{min}$. This makes sense given that more negative $\lambda$ is associated with more events with low X-ray fluence, so raising the minimum fluence is necessary to retain the balance between events with and without detectable signal.

The CDFs computed from the best fit parameters are also shown in Figure 3 as thin curves, but they overlap with the data so much that no systematic deviations are evident. A residuals plot showing the differences between the two sets of curves is shown in Figure 4. Some deviations can be seen but are well within the typical size of the fluctuations due to random distribution of points in the data set.

As a sanity check, Monte Carlo simulations of data sets drawn from the distributions described above show no obvious deviations from the distribution of the data when plotted as in Figure 2. It is also worth noting that since our fitting procedure captures the joint distribution, it also captures the distribution within each detector separately. In this paper, we have not attempted to compensate for saturation to construct a true energy distribution, but *Kochkin et al.* [2015], using essentially the same experimental setup as used here, do present such a spectrum [*Kochkin et al.*, 2015, Figure 12] constructed by a sophisticated procedure of fitting pulse shapes to observed oscilloscope records. Though the analysis in *Kochkin et al.* [2015] includes the data set used here, the analysis procedures are completely different, and *Kochkin et al.* [2015] include an additional 2000 sparks and thus represent an approximately independent analysis. The *Kochkin et al.* [2015] results are presented for energy deposition per burst not per spark, but the fact that relatively few sparks have multiple bursts means that it is still useful to directly compare the results from *Kochkin et al.* [2015] to equation (5). Evaluation of equation (5) with our fit result values of $\lambda$, $\mu$, $\eta_{min}$, and $\eta_{max}$, normalized and overlaid with the data from *Kochkin et al.* [2015], is shown in Figure 5, demonstrating much better agreement than the single exponential distribution used in *Kochkin et al.* [2015]. *Kochkin et al.* [2015] inferred a 200 keV characteristic burst energy (which is asserted to be roughly equal to characteristic photon energy) but showed an extremely poor fit at high burst energies that they attribute to pileup. Properly accounting for pileup as we do here with the Poisson distribution of observed counts ($p_\eta(N)$) and for fluence variability with the assumption of a power law ($p_\lambda(\eta)$) gives us a much better fit with a much lower mean photon energy (86 keV), suggesting that pileup and fluence variation is essential to consider in interpretation of such data.

As a final comparison to data, it is useful to consider attenuator experiments. While we ran no attenuator experiments in the data set described above, *Kochkin et al.* [2015] describe such experiments, again with essentially the same setup as used here. They report results for lead attenuators of varying thickness, determining the fraction of sparks that produce observable signals in a single detector by running 50 sparks with each attenuator.

The distributions determined above can easily be used to construct results for attenuated scenarios for comparison with the observations in *Kochkin et al.* [2015] by Monte Carlo simulation. First, we draw a random spark fluence, then draw a random number of photons from such a spark, then draw that many random photon energies from the photon energy spectrum, then attenuate those photons probabilistically by use of the mass attenuation coefficients from *Hubbell and Seltzer* [2004], and then finally classify the event as containing detected X-rays or not based on whether or not any photons made it through the lead shield. We repeat this procedure $10^6$ times, compute the overall fraction of events with detected X-rays, and repeat for each attenuator thickness. Results are shown in Table 1. Comparison with data requires an estimate of the uncertainty in the data, here estimated by calculating a 68% confidence interval based on the binomial distribution. Sixty-eight percent confidence corresponds to a $\pm 1\sigma$ error bar, and all but one of our predictions are consistent with the observations at this confidence level. The distributions determined in this paper are thus also consistent with attenuator experiments, though further attenuator observations would be useful to narrow the uncertainties in the observations.

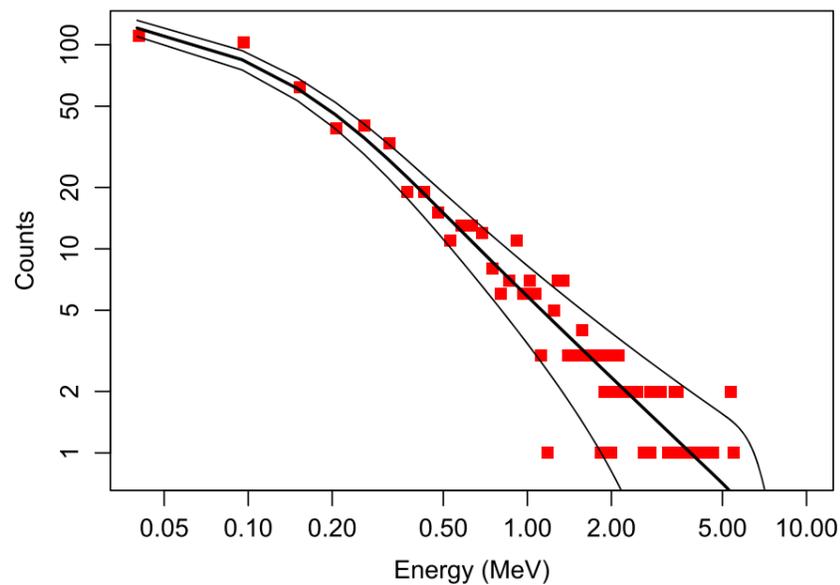

**Figure 5.** Comparison of the predicted total energy deposition spectrum from this paper with observations from Figure 12 of *Kochkin et al.* [2015]. The red squares show the observed spectrum, while the black curves show the predictions based on our distributions. The thick curve is the prediction, while the thin curves represent the $\pm\sqrt{N}$ standard deviation expected for the spread in the data.

## 5. Discussion

To summarize, we have studied observations of X-ray emission by sparks and attempted to understand those observations by modeling the process as a combination of a distribution of spark fluence, Poisson statistics, and an X-ray energy spectrum. These distributions successfully reproduce not only the data used here but also independent analysis of similar experiments on overall energy distribution and signal attenuation in *Kochkin et al.* [2015]. The two main results are a well-defined photon mean energy and the distribution of overall spark fluence.

The $86 \pm 7$ keV mean photon energy determined above is consistent with most earlier estimates but somewhat lower than some. We feel this mean energy is quite reasonable given that electrons accelerated in the environment of a 1 MV spark realistically must have at most a few hundred keV of energy and bremsstrahlung photons produced must have a lower energy than that of the source electrons. As a crude estimate of electron energy, assume bremsstrahlung produced by the relatively low energy electrons here follows approximately the $dN/dE_\gamma \propto 1/E_\gamma$ distribution seen for bremsstrahlung produced by high-energy electrons. This $1/E_\gamma$ distribution spans the range from the energy of the electron as a maximum to a minimum energy at most 20 keV. Using 20 keV as the minimum energy and requiring that the average photon energy is 86 keV require the electron energy to be roughly 200 keV, while using 5 keV as the minimum energy gives nearly 400 keV as the electron energy. This 200–400 keV energy range is not inconsistent with the potential available for electron acceleration at the time of X-ray production as seen in Figure 1, but this is a crude estimate in need of refinement by modeling of electron energy distributions and the resulting bremsstrahlung before it can be taken very seriously.

**Table 1.** Attenuator Experiments From *Kochkin et al.* [2015] and Predictions Based on the Distributions Determined in This Paper[a]

| Thickness (mm) | Observed Fraction (%) | 68% Confidence Interval (%) | Predicted Fraction (%) |
|---|---|---|---|
| 0.0 | 32 | 25–40 | 35 |
| 1.5 | 8 | 4–14 | 11 |
| 3.0 | 5 | 2.7–11 | 6.6 |
| 4.5 | 0 | 0–3.6 | 4.2 |
| 6.0 | 0 | 0–3.6 | 2.8 |
| 7.5 | 2 | 0.3–6.4 | 1.9 |

[a]The 68% confidence intervals correspond to $\pm 1\sigma$ error bars and are calculated based on binomial statistics.

The exponential X-ray spectrum motivated by the expectation of an at least approximately exponential distribution in electron energies very closely reproduces the observed distributions, though is likely not the only such distribution to do so, and as noted above we have assumed that fluence ($\eta$) and mean energy are independent which may or may not be true. While the average energy should be close to correct, the other distributions at work in this system complicate the analysis.

The more important factor at work is the very broad distribution of observed spark X-ray fluence ($\eta$), which, as mentioned earlier, we assume captures both intrinsic variability in spark luminosity and variability in the spatial distribution of X-ray emissions. The results obtained above, which the distribution of $\eta$ is very hard (power law index −1.29) and covers a broad range, from 0.02 photons to over 100 photons for the experiment conducted here, pose a challenge to the idea that the strong field near the head of a negative streamer is all that is required to accelerate electrons into the energy regime needed for 100 keV X-ray production. All sparks involve extensive streamer production [see, e.g., *Kochkin et al.*, 2014, Figure 2], so the number or distribution of streamer production cannot directly explain the broad distribution of spark X-ray fluence. As discussed in *Kochkin et al.* [2015], the very short duration and appearance of multiple bursts of X-ray emissions suggests that X-ray emission occurs during some fast transient process like streamer collision, but it is not clear how streamer collision frequency would be distributed and whether such a distribution could reproduce the characteristics observed here. While we have not proved that X-ray fluence truly follows a power law, only that our assumption of a power law is consistent with our data, it is perhaps not unreasonable that a power law distribution might arise here in the context of dielectric breakdown since power laws arise in studies of systems that similarly approach breakdown and then collapse [*Bak et al.*, 1987]. Regardless of the true form of the distribution, we hope our analysis here is useful for evaluation of X-ray production mechanisms, which must explain the frequency of both dim and bright sparks in roughly the balance described here.

As for the relative contributions of spark luminosity variation and nonuniform X-ray spatial distribution to the observed fluence variability, some crude analysis is instructive. Assuming no intrinsic luminosity variability, the observed fluence variability should correspond roughly to the degree of spatial variability in X-ray emissions. One can set an upper limit on the spatial variability present in X-ray emissions in an experiment such as this by examining the directional distribution of bremsstrahlung from a unidirectional beam of 500 keV electrons. Such directional distributions can be found, for example, in *Tseng et al.* [1979, Figure 5] and *Köhn and Ebert* [2014, Figure 7b]. Both figures present plots of the intensity of emissions versus angle relative to the direction of the electron beam, with *Tseng et al.* [1979] showing a factor of around 100 difference from highest to lowest fluence, while *Köhn and Ebert* [2014] show a factor of around 300. That the best fit distribution constructed here requires a factor of $\eta_{max}/\eta_{min} = 100/0.02 = 5000$ from highest to lowest fluence suggests that even this extremely aggressive assumption of unidirectional 500 keV electrons falls short of explaining the observed fluence variability, but this analysis is admittedly crude.

A slightly less crude assessment can be made by Monte Carlo: first, draw a random beam direction uniformly from the high-voltage electrode in a hemisphere facing the ground electrode, then determine the angle from the center of that beam to the location of the detector and evaluate the relative fluence by extracting the distributions shown in *Tseng et al.* [1979] and *Köhn and Ebert* [2014] for a relatively high energy photon. Though this ignores the X-ray energy dependence of the bremsstrahlung angular distribution, it should reasonably account for the degree of variability. Repeat this procedure many times, compiling a histogram estimating the fluence distribution, then scale these histograms such that they are normalized and adjust the relative fluence to correspond to $\eta$ by shifting the relative fluence until the histograms correctly predict the fraction of events that would fall at intensities too low to be detected. Histograms constructed in this manner are compared to the best fit $\eta$ power law in Figure 6.

The similarity of slopes is striking, especially for the *Köhn and Ebert* [2014] case; but it is clear that fluence variability due to beaming cannot fully account for the observations. In the context of the 86 keV average photon energy determined here, the maximum fluence of the bremsstrahlung-derived cases, $\eta \sim 10$, would result in 800 keV deposited on average, far from the 3–5 MeV needed to saturate the detectors. Put another way, if we attempt to account for the large fraction of low-fluence events simply as emissions beamed away from the detectors, the maximum fluence would then be too low to account for events that saturate the detectors. The scale of variability in the distributions derived from bremsstrahlung, even for the extreme case of unidirectional 500 keV electrons, is therefore too low to fully account for the variability observed, providing clear evidence for strong spark-to-spark variability. Crudely comparing the spread of the *Köhn and Ebert* [2014] curve to our power law in Figure 6, there is at least another order of magnitude variability in fluence that cannot be explained as spatial variability and that is a very conservative estimate since in a more realistic scenario there will be directional dispersion of electrons, lower energy electrons, and possibly multiple beams, all of which lead to reductions in the spatial variability due to beaming alone, thus requiring even more spark-to-spark variation.

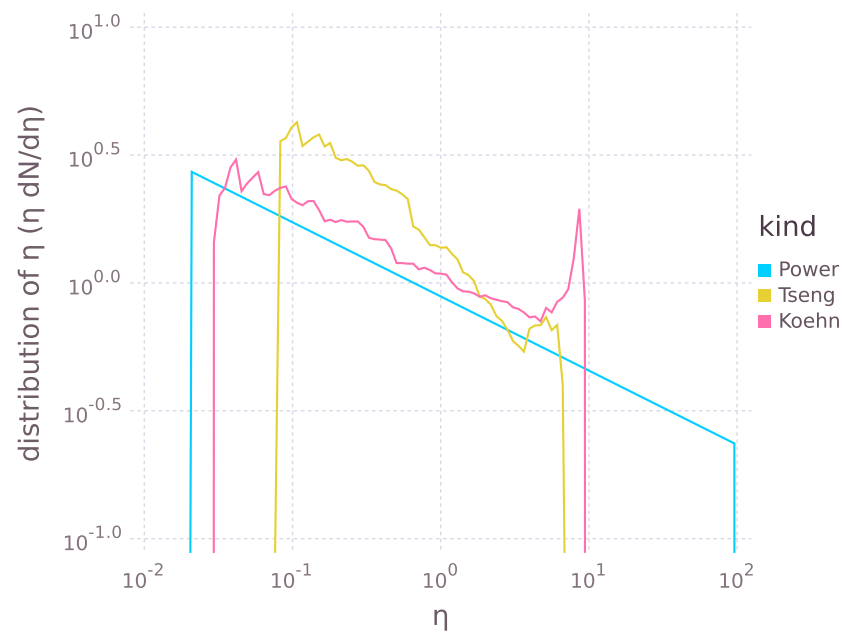

**Figure 6.** Predicted fluence distributions assuming only spatial variability due to beaming of bremsstrahlung from unidirectional 500 keV electrons. The curve marked "Tseng" is derived from Figure 5 of *Tseng et al.* [1979], "Koehn" is derived from Figure 7b of *Köhn and Ebert* [2014], and "Power" is the best fit power law determined here.

We hope the results described above will provide a basis for comparison to results of other spark experiments to determine if the distributions of mean photon energy and spark X-ray fluence depend on the experimental setup. We further hope that theoretical work can find a basis for such distributions to confirm or refute our assumptions and that such theories can shed light on the process of runaway electron acceleration in sparks as suggested to be relevant to other electrical phenomena such as lightning or terrestrial gamma ray flashes.

**Acknowledgments**
The data described in this paper are available from the authors on request (bcarlson1@carthage.edu). We are very grateful to Lex van Deursen for the valuable discussions. This study was supported by the European Research Council under the European Union's Seventh Framework Programme (FP7/2007-2013)/ERC grant agreement 320839 and the Research Council of Norway under contracts 208028/F50, 216872/F50, and 223252/F50 (CoE) and by the Carthage Summer Undergraduate Research Experience program.

**References**

Bak, P., C. Tang, and K. Wiesenfeld (1987), Self-organized criticality: An explanation of the 1/f noise, *Phys. Rev. Lett.*, *59*(4), 381–384, doi:10.1103/PhysRevLett.59.381.
Carlson, B. E. (2009), Terrestrial gamma-ray flash production by lightning, PhD thesis, Stanford Univ., Stanford, Calif.
Carlson, B. E., N. G. Lehtinen, and U. S. Inan (2008), Runaway relativistic electron avalanche seeding in the Earth's atmosphere, *J. Geophys. Res.*, *113*, A10307, doi:10.1029/2008JA013210.
Chanrion, O., and T. Neubert (2010), Production of runaway electrons by negative streamer discharges, *J. Geophys. Res.*, *115*, A00E32, doi:10.1029/2009JA014774.
Dwyer, J. R. (2003), A fundamental limit on electric fields in air, *Geophys. Res. Lett.*, *30*(20), 2055, doi:10.1029/2003GL017781.
Dwyer, J. R. (2005a), X-ray bursts produced by laboratory sparks in air, *Geophys. Res. Lett.*, *32*, L20809, doi:10.1029/2005GL024027.
Dwyer, J. R. (2005b), X-ray bursts associated with leader steps in cloud-to-ground lightning, *Geophys. Res. Lett.*, *32*, L01803, doi:10.1029/2004GL021782.
Dwyer, J. R. (2007), Relativistic breakdown in planetary atmospheres, *Phys. Plasmas*, *14*(4), 42901, doi:10.1063/1.2709652.
Dwyer, J. R., and M. A. Uman (2014), The physics of lightning, *Phys. Rep.*, *534*(4), 147–241, doi:10.1016/j.physrep.2013.09.004.
Dwyer, J. R., Z. Saleh, H. K. Rassoul, D. Concha, M. Rahman, V. Cooray, J. Jerauld, M. A. Uman, and V. A. Rakov (2008), A study of X-ray emission from laboratory sparks in air at atmospheric pressure, *J. Geophys. Res.*, *113*, D23207, doi:10.1029/2008JD010315.
Dwyer, J. R., M. Schaal, H. K. Rassoul, M. A. A. Uman, D. M. Jordan, and D. Hill (2011), High-speed X-ray images of triggered lightning dart leaders, *J. Geophys. Res.*, *116*, D20208, doi:10.1029/2011JD015973.
Dwyer, J. R., et al. (2003), Energetic radiation produced during rocket-triggered lightning, *Science*, *299*(5607), 694–697, doi:10.1126/science.1078940.
Fasano, G., and A. Franceschini (1987), A multidimensional version of the Kolmogorov-Smirnov test, *Mon. Not. R. Astron. Soc.*, *225*, 155–170.
Gurevich, A., G. Milikh, and R. Roussel-Dupre (1992), Runaway electron mechanism of air breakdown and preconditioning during a thunderstorm, *Phys. Lett. A*, *165*(5–6), 463–468, doi:10.1016/0375-9601(92)90348-P.
Howard, J., M. A. Uman, C. Biagi, D. Hill, J. Jerauld, V. A. Rakov, J. Dwyer, Z. Saleh, and H. Rassoul (2010), RF and X-ray source locations during the lightning attachment process, *J. Geophys. Res.*, *115*, D06204, doi:10.1029/2009JD012055.
Hubbell, J. H., and S. M. Seltzer (2004), Tables of X-ray mass attenuation coefficients and mass energy-absorption coefficients from 1 keV to 20 MeV for elements Z = 1 to 92 and 48 additional substances of dosimetric interest, Version 1.4, Report NISTIR-5632, (National Institute of Standards and Technology, 1995). [Available at http://physics.nist.gov/xaamdi.]
Johnson, S. J. (2014), The NLopt nonlinear-optimization package.
Kochkin, P. O., C. V. Nguyen, A. P. J. van Deursen, and U. Ebert (2012), Experimental study of hard X-rays emitted from metre-scale positive discharges in air, *J. Phys. D: Appl. Phys.*, *45*(42), 425202, doi:10.1088/0022-3727/45/42/425202.
Kochkin, P. O., A. P. J. van Deursen, and U. Ebert (2014), Experimental study of the spatio-temporal development of metre-scale negative discharge in air, *J. Phys. D: Appl. Phys.*, *47*(14), 145203, doi:10.1088/0022-3727/47/14/145203.

Kochkin, P. O., A. P. J. van Deursen, and U. Ebert (2015), Experimental study on hard X-rays emitted from metre-scale negative discharges in air, *J. Phys. D: Appl. Phys.*, *48*(2), 025205, doi:10.1088/0022-3727/48/2/025205.

Köhn, C., and U. Ebert (2014), Angular distribution of Bremsstrahlung photons and of positrons for calculations of terrestrial gamma-ray flashes and positron beams, *Atmos. Res.*, *135-136*, 432–465, doi:10.1016/j.atmosres.2013.03.012.

Li, C., U. Ebert, and W. Hundsdorfer (2009), 3D hybrid computations for streamer discharges and production of runaway electrons, *J. Phys. D: Appl. Phys.*, *42*(20), 202003, doi:10.1088/0022-3727/42/20/202003.

March, V., and J. Montanyà (2010), Influence of the voltage-time derivative in X-ray emission from laboratory sparks, *Geophys. Res. Lett.*, *37*, L19801, doi:10.1029/2010GL044543.

March, V., and J. Montanyà (2011), X-rays from laboratory sparks in air: The role of the cathode in the production of runaway electrons, *Geophys. Res. Lett.*, *38*, L04803, doi:10.1029/2010GL046540.

Moore, C. B., K. B. Eack, G. D. Aulich, and W. Rison (2001), Energetic radiation associated with lightning stepped leaders, *Geophys. Res. Lett.*, *28*(11), 2141–2144, doi:10.1029/2001GL013140.

Moss, G. D., V. P. Pasko, N. Liu, and G. Veronis (2006), Monte Carlo model for analysis of thermal runaway electrons in streamer tips in transient luminous events and streamer zones of lightning leaders, *J. Geophys. Res.*, *111*, A02307, doi:10.1029/2005JA011350.

Nelder, J. A., and R. Mead (1965), A simplex method for function minimization, *Comput. J.*, *7*(4), 308–313, doi:10.1093/comjnl/7.4.308.

Nguyen, C. V. (2012), Experimental study on hard radiation from long laboratory spark discharges in air, PhD thesis, Technische Universiteit Eindhoven, Netherlands, doi:10.6100/IR731153.

Nguyen, C. V., A. P. J. van Deursen, and U. Ebert (2008), Multiple X-ray bursts from long discharges in air, *J. Phys. D: Appl. Phys.*, *41*(23), 234012, doi:10.1088/0022-3727/41/23/234012.

Ostgaard, N., B. E. Carlson, O. Grondahl, P. Kochkin, R. Nisi, and T. Gjesteland (2014), Search for X-rays and relativistic electrons in laboratory discharge experiments, *Presented at the American Geophysical Union Fall Meeting, Wien, Austria*.

Rahman, M., V. Cooray, N. A. Ahmad, J. Nyberg, V. A. Rakov, and S. Sharma (2008), X rays from 80-cm long sparks in air, *Geophys. Res. Lett.*, *35*, L06805, doi:10.1029/2007GL032678.

Razali, N., and Y. Wah (2011), Power comparisons of Shapiro-Wilk, Kolmogorov-Smirnov, Lilliefors and Anderson-Darling tests, *J. Stat. Model. Anal.*, *2*(1), 21–33.

Tseng, H. K., R. H. Pratt, and C. M. Lee (1979), Electron bremsstrahlung angular distributions in the 1–500 keV energy range, *Phys. Rev. A*, *19*(1), 187–195, doi:10.1103/PhysRevA.19.187.

Wilson, C. T. R. (1925), The acceleration of $\beta$-particles in strong electric fields such as those of thunderclouds, *Math. Proc. Cambridge Philos. Soc.*, *22*(04), 534–538, doi:10.1017/S0305004100003236.